\documentclass{article}
\usepackage[linesnumbered,ruled,vlined]{algorithm2e}
\usepackage{color, colortbl}
\usepackage{booktabs} % for professional tables
\usepackage{multirow}
\usepackage{wrapfig}
\usepackage{appendix}
\usepackage{amsmath}
\usepackage{tcolorbox}
    \usepackage[preprint,nonatbib]{neurips_2024}

\usepackage[utf8]{inputenc} % allow utf-8 input
\usepackage[T1]{fontenc}    % use 8-bit T1 fonts
\usepackage{url}            % simple URL typesetting
\usepackage{booktabs}       % professional-quality tables
\usepackage{amsfonts}       % blackboard math symbols
\usepackage{nicefrac}       % compact symbols for 1/2, etc.
\usepackage{microtype}      % microtypography
\usepackage{xcolor}         % colors
\usepackage{float}
\usepackage{graphicx}
\usepackage{hyperref}       % hyperlinks
\usepackage{caption}
\usepackage{placeins} 
\usepackage{lipsum}  
\usepackage{float}
\usepackage{listings}

\lstdefinestyle{yaml}{
    basicstyle=\ttfamily\footnotesize,
    commentstyle=\color{gray}\itshape,
    stringstyle=\color{green!60!black},
    keywordstyle=\color{black},
    morekeywords={true,false,null,yes,no},
    numbers=left,
    numberstyle=\tiny\color{gray},
    stepnumber=1,
    numbersep=5pt,
    backgroundcolor=\color{white},
    showspaces=false,
    showstringspaces=false,
    showtabs=false,
    frame=single,
    rulecolor=\color{black},
    tabsize=2,
    breaklines=true,
    breakatwhitespace=true,
    captionpos=b,
    literate={*}{{\char42}}1 {-}{{\char45}}1 {:}{{\char58}}1
}
\usepackage{framed}
\title{NeuGaze: Reshaping the future BCI}

\author{\parbox{13cm}
  {\centering
    {Yiqian Yang$^{1}$
  }
}
}
\begin{document}
\maketitle
\footnotetext[1]{The Hong Kong University of Science and Technology (Guangzhou), People's Republic of China}

\begin{abstract}

Traditional brain-computer interfaces (BCIs), reliant on costly electroencephalography or invasive implants, struggle with complex human-computer interactions due to setup complexity and limited precision. We present NeuGaze, a novel webcam-based system that leverages eye gaze, head movements, and facial expressions to enable intuitive, real-time control using only a standard 30 Hz webcam, often pre-installed in laptops. Requiring minimal calibration, NeuGaze achieves performance comparable to conventional inputs, supporting precise cursor navigation, key triggering via an efficient skill wheel, and dynamic gaming interactions, such as defeating formidable opponents in first-person games. By harnessing preserved neck-up functionalities in motor-impaired individuals, NeuGaze eliminates the need for specialized hardware, offering a low-cost, accessible alternative to BCIs. This paradigm empowers diverse applications, from assistive technology to entertainment, redefining human-computer interaction for motor-impaired users. Project is at \href{https://github.com/NeuSpeech/NeuGaze}{github.com/NeuSpeech/NeuGaze}.

\end{abstract}

\section{Introduction}

Brain-computer interfaces (BCIs) have emerged as transformative tools, bridging the gap between human cognition and computational systems. By translating neural activity into actionable commands, BCIs hold immense potential for applications ranging from assistive technologies to entertainment and beyond. However, achieving seamless, complex human-computer interaction remains a significant challenge for existing BCI approaches. Current practical BCIs primarily fall into two categories: non-invasive systems, such as those utilizing electroencephalography (EEG) within the visual evoked potential (VEP) paradigm, and invasive systems, which rely on implanted electrodes to capture neural signals directly.

\subsection{Invasive Devices}

Invasive brain-computer interfaces (BCIs) involve surgically implanting electrodes into brain tissue to record high-resolution neural signals, enabling precise control of external devices. This section explores their historical development, operational principles, key technologies, target users, market trends, and challenges.

\textbf{Background}: The concept of invasive BCIs emerged in the late 1990s with experiments demonstrating neural control of robotic devices in primates \cite{chapin1999real}. Advances in electrode technology and neural decoding algorithms in the 2000s enabled human applications, particularly for individuals with severe motor impairments \cite{hochberg2006neuronal}. Companies like Neuralink and Synchron have since advanced the field, moving from research to clinical trials \cite{musk2019neuralink, oxley2016minimally}.

\textbf{Operational Principles}: Invasive BCIs use microelectrode arrays implanted in the cortex to capture action potentials or local field potentials \cite{lebedev2017brain}. Signal processing employs machine learning, such as neural networks, to decode motor intentions with low latency \cite{anumanchipalli2019speech}. Implantation typically requires craniotomy or endovascular delivery, using biocompatible materials like platinum \cite{polikov2005response}. Challenges include tissue scarring and signal degradation over time, which can reduce long-term efficacy \cite{polikov2005response}.

\textbf{Key Technologies}: \textbf{Neuralink} develops the Telepathy N1 implant, with thousands of electrodes to record neural activity at high frequency \cite{musk2019neuralink}. A robotic system inserts ultra-thin electrodes to minimize tissue damage, and early human trials in 2024 enabled cursor control for quadriplegic patients \cite{neuralink2024first}. \textbf{Synchron}’s Stentrode, delivered endovascularly, uses a stent with electrodes to record motor cortex signals, supporting digital device control for ALS patients \cite{oxley2016minimally, oxley2021endovascular}. \textbf{Neuroxess} focuses on high-density microelectrode arrays, achieving single-neuron resolution in preclinical trials \cite{neuroxess2023highdensity}.

\textbf{Target Users}: Invasive BCIs primarily serve individuals with severe motor impairments, such as quadriplegia, ALS, and locked-in syndrome. These systems enable tasks like robotic arm control, cursor navigation, and communication, restoring independence for patients with intact cognitive function \cite{hochberg2012reach, oxley2021endovascular}.

\textbf{Market Trends}: The BCI market, including invasive systems, is growing due to increasing prevalence of neurological disorders and advancements in neurotechnology \cite{grandview2023bci}. North America leads in adoption, driven by robust healthcare infrastructure \cite{grandview2023bci}. High costs and regulatory hurdles, however, limit broader access \cite{synchron2021fda}.

\textbf{Challenges and Future Trends}: Surgical risks, biocompatibility issues, and signal degradation pose challenges \cite{polikov2005response}. Ethical concerns, including neural privacy, also arise \cite{musk2019neuralink}. Future developments include biodegradable electrodes, wireless power systems, and AI-enhanced decoding to improve reliability \cite{lebedev2017brain}. Hybrid systems combining invasive and non-invasive approaches are also emerging \cite{grandview2023bci}.

\textbf{Summary}: Invasive BCIs offer high-fidelity neural control for motor-impaired individuals but face surgical and ethical challenges. Ongoing innovations in materials and algorithms will enhance their clinical impact.

\subsection{Non-Invasive Devices}

Non-invasive BCIs use scalp-based sensors to measure brain activity, offering a safer alternative to invasive systems. This section focuses on Visual Evoked Potential (VEP)-based BCIs, covering their background, principles, key products, users, market trends, and limitations.

\textbf{Background}: Non-invasive BCIs originated in the 1970s with electroencephalography (EEG) for brain signal detection \cite{vidal1973toward}. Steady-state VEP (SSVEP) systems, leveraging visual stimuli to elicit neural responses, became prominent in the 1990s due to their robustness \cite{vialatte2010steady}. These systems have evolved into commercial products for assistive and consumer applications \cite{allison2010toward}.

\textbf{Operational Principles}: SSVEP-BCIs use flickering visual stimuli (5-20 Hz) to generate frequency-locked responses in the occipital cortex, detected via EEG electrodes \cite{vialatte2010steady}. Signal processing, such as Canonical Correlation Analysis, identifies the target frequency with high accuracy \cite{zhu2010design}. Noise from artifacts requires filtering, and multi-channel setups improve resolution but increase complexity \cite{vialatte2010steady}. Dry electrodes simplify use but reduce signal quality \cite{zhu2010design}.

\textbf{Key Products}: \textbf{g.tec}’s Intendix system supports SSVEP-based spelling and navigation for locked-in patients \cite{guger2012intendix}. \textbf{NeuroSky}’s MindWave, an affordable single-channel headset, enables attention-based gaming \cite{neurosky2023mindwave}. \textbf{Emotiv}’s EPOC+ uses multiple channels for SSVEP applications in virtual reality and neurofeedback \cite{emotiv2023epoc}. These products balance cost and functionality for different markets \cite{allison2010toward}.

\textbf{Target Users}: VEP-BCIs benefit patients with ALS, quadriplegia, and locked-in syndrome, enabling communication and device control \cite{wolpaw2002brain}. Stroke survivors also use these systems for assistive tasks, though efficacy varies \cite{zhu2010design}. They are less suitable for visually impaired or photosensitive individuals \cite{vialatte2010steady}.

\textbf{Market Trends}: Non-invasive BCIs dominate the market due to their accessibility, with growth driven by neurological disorder prevalence and consumer demand for gaming and neurofeedback \cite{grandview2023bci}. North America and Europe lead in adoption \cite{grandview2023bci}.

\textbf{Challenges and Future Trends}: Low signal-to-noise ratios and user fatigue limit performance \cite{wolpaw2002brain}. Environmental factors also affect accuracy \cite{vialatte2010steady}. Future trends include hybrid BCIs, improved dry electrodes, and AI-based signal processing to enhance usability \cite{allison2010toward}. Integration with virtual reality could expand applications \cite{grandview2023bci}.

\textbf{Summary}: VEP-BCIs provide accessible solutions for motor-impaired users but are constrained by signal quality and fatigue. Advances in hardware and AI will broaden their impact.

\subsection{Applicable Medical Conditions}

Brain-computer interfaces (BCIs) and assistive technologies provide critical solutions for individuals with severe motor impairments, enabling communication and control when traditional methods are unavailable. This section examines four medical conditions—high-level quadriplegia, amyotrophic lateral sclerosis (ALS), severe post-stroke motor impairments, and congenital limb deformities (e.g., phocomelia)—detailing their pathophysiology, functional body parts above the neck, applicable BCI or assistive devices, challenges, and market relevance. A summary highlights a key insight: the preserved functionality of neck-up regions in these populations, suggesting alternative non-BCI approaches.

\textbf{High-Level Quadriplegia}:

\textit{Pathophysiology}: High-level quadriplegia results from cervical spinal cord injuries (C1-C4), disrupting motor and sensory pathways below the neck. These injuries often stem from trauma, affecting motor neurons and leaving patients with no limb or trunk control \cite{nscisc2023spinal}.

\textit{Functional Body Parts}: Cognitive functions, eye movements, facial muscles, and head movements remain intact in most cases, enabling gaze tracking, facial expression recognition, and limited head-based control \cite{hochberg2012reach}.

\textit{BCI or Assistive Devices}: Invasive BCIs, such as Neuralink’s Telepathy N1, use microelectrode arrays to decode motor cortex signals for robotic arm or cursor control \cite{musk2019neuralink}. Non-invasive Visual Evoked Potential (VEP)-based BCIs, like g.tec’s Intendix, leverage EEG to enable wheelchair navigation or text input via eye gaze \cite{guger2012intendix}. Assistive devices, such as eye-tracking systems (e.g., Tobii Dynavox), support communication through gaze-based interfaces \cite{zhu2010design}.

\textit{Challenges}: Invasive BCIs carry surgical risks (e.g., infection, tissue damage) and high costs, limiting accessibility \cite{hochberg2012reach}. Non-invasive systems suffer from low signal resolution, requiring extended training (weeks) and causing user fatigue \cite{zhu2010design}. Eye-tracking devices depend on consistent lighting and user calibration \cite{zhu2010design}.

\textit{Market}: Quadriplegia drives significant BCI demand, contributing to the medical BCI market’s growth, valued at \$1.5 billion in 2022 \cite{grandview2023bci}. North America leads due to advanced healthcare infrastructure \cite{grandview2023bci}.

\textbf{Amyotrophic Lateral Sclerosis (ALS)}:

\textit{Pathophysiology}: ALS is a progressive neurodegenerative disease causing motor neuron loss, leading to paralysis within 3-5 years. It spares cognitive and sensory functions but results in locked-in states in advanced stages \cite{rowland2001amyotrophic}.

\textit{Functional Body Parts}: Early-stage ALS patients retain head, facial, and eye movement capabilities. In locked-in states, eye movements and subtle facial muscle activity (e.g., blinks) often persist, supporting gaze- or expression-based control \cite{guger2012intendix}.

\textit{BCI or Assistive Devices}: Minimally invasive BCIs, like Synchron’s Stentrode, record motor cortex signals endovascularly to control digital devices \cite{oxley2021endovascular}. Non-invasive VEP-BCIs, such as the Intendix system, enable communication via EEG-based gaze detection \cite{guger2012intendix}. Assistive eye-tracking devices (e.g., EyeSpeak) facilitate text input using blinks or gaze \cite{zhu2010design}.

\textit{Challenges}: Neural degeneration in ALS requires frequent BCI recalibration, and non-invasive systems face signal degradation from muscle artifacts \cite{guger2012intendix}. Assistive devices are limited by cost and setup complexity, particularly in home settings \cite{zhu2010design}.

\textit{Market}: ALS contributes to the growing BCI market, with demand driven by the need for communication tools in locked-in patients. The market’s expansion is supported by increasing neurological disorder prevalence \cite{grandview2023bci}.

\textbf{Severe Post-Stroke Motor Impairments}:

\textit{Pathophysiology}: Stroke results from cortical or subcortical lesions, causing hemiplegia or severe motor deficits in limbs. Cognitive and sensory functions are often preserved, though lesion variability affects motor outcomes \cite{who2023stroke}.

\textit{Functional Body Parts}: Head movements, facial expressions, and eye gaze remain functional in most stroke patients, enabling control via gaze tracking or head-mounted sensors \cite{biasiucci2018brain}.

\textit{BCI or Assistive Devices}: Invasive BCIs with functional electrical stimulation (FES) use cortical implants to trigger muscle contractions for rehabilitation \cite{biasiucci2018brain}. Non-invasive EEG-based BCIs, using motor imagery or SSVEP, support smart home control or prosthetic operation \cite{zhu2010design}. Assistive head-tracking devices (e.g., HeadMouse Nano) allow cursor control via head movements \cite{zhu2010design}.

\textit{Challenges}: Lesion heterogeneity necessitates personalized BCI algorithms, increasing development costs \cite{biasiucci2018brain}. Non-invasive systems have lower accuracy in noisy environments, and assistive devices require user training to achieve proficiency \cite{zhu2010design}.

\textit{Market}: Stroke is a major driver of BCI demand, with applications in rehabilitation and assistive control fueling market growth, particularly in North America and Europe \cite{grandview2023bci}.

\textbf{Congenital Limb Deformities (e.g., Phocomelia)}:

\textit{Pathophysiology}: Phocomelia involves partial or absent limb development due to genetic or environmental factors, leaving motor functions severely restricted while cognitive and sensory capabilities are intact \cite{cdc2023congenital}.

\textit{Functional Body Parts}: Head, facial, and eye movements are fully functional, supporting gaze-based or head-motion-based interfaces for control \cite{zhu2010design}.

\textit{BCI or Assistive Devices}: Non-invasive VEP-BCIs enable prosthetic control or computer access via EEG-based gaze detection \cite{zhu2010design}. Assistive devices, such as eye-tracking systems or head-mounted controllers (e.g., GlassOuse), allow hands-free operation of computers or prosthetics \cite{zhu2010design}. Invasive BCIs are rarely used due to surgical risks in younger patients \cite{lebedev2017brain}.

\textit{Challenges}: Non-invasive BCIs have limited signal resolution, and assistive devices face durability issues (e.g., wear on headsets) \cite{zhu2010design}. Costs and accessibility remain barriers, particularly for pediatric populations \cite{grandview2023bci}.

\textit{Market}: Congenital deformities represent a smaller but growing segment of the BCI market, driven by demand for pediatric assistive technologies \cite{grandview2023bci}.

% \textbf{Summary and Proposed Insight}: BCIs and assistive devices significantly enhance communication and control for individuals with quadriplegia, ALS, stroke, and congenital deformities, addressing severe motor impairments where cognitive function is preserved. Invasive BCIs offer high precision but face surgical and cost barriers, while non-invasive systems and assistive devices provide accessibility at the expense of signal quality \cite{hochberg2012reach, zhu2010design}. The medical BCI market is expanding, driven by these conditions, with North America leading adoption \cite{grandview2023bci}. However, challenges such as high costs, training requirements, and environmental limitations persist across all conditions \cite{lebedev2017brain}.

% A critical observation from this analysis is that most BCI target populations retain functional capabilities above the neck, including head movements, facial expressions, and eye gaze. These preserved functions are already leveraged by assistive devices like eye-tracking and head-mounted systems, suggesting that traditional BCIs may not be the only or optimal solution \cite{zhu2010design}. We propose exploring alternative methods that directly utilize head, facial, and eye-gaze movements to create low-cost, accessible interfaces. Such approaches could bypass the complexity and cost of neural signal processing, offering a practical solution for these populations while addressing the same functional needs as BCIs.

Brain-computer interfaces (BCIs) and assistive devices significantly enhance communication and control for individuals with severe motor impairments, such as quadriplegia, ALS, stroke, and congenital deformities, where cognitive functions remain intact \cite{hochberg2012reach}. However, invasive BCIs, while precise, face surgical risks and high costs, and non-invasive systems, like EEG-based visual evoked potentials, suffer from low signal quality and user fatigue \cite{zhu2010design, lebedev2017brain}. These limitations hinder their ability to support complex, real-time human-computer interactions, particularly for motor-impaired users seeking accessible solutions \cite{grandview2023bci}.

A critical insight from this analysis is that most BCI target populations retain functional capabilities above the neck, including eye gaze, head movements, and facial expressions, which are underutilized in traditional BCIs \cite{zhu2010design}. We propose NeuGaze, a novel webcam-based framework that leverages these preserved functionalities to enable intuitive, hands-free control without specialized hardware. Using only a standard 30 Hz webcam, NeuGaze integrates gaze estimation, head pose tracking, and facial expression detection to achieve performance comparable to conventional inputs. Its key contributions include: (1) a low-cost, accessible design requiring minimal calibration, (2) real-time processing for dynamic tasks like first-person gaming, and (3) an efficient skill wheel mechanism that reduces the number of facial expressions needed for complex key mappings. This paper introduces NeuGaze’s methodology, evaluates its performance in controlling a challenging action game, and discusses its potential to transform assistive technology and human-computer interaction.

\section{Methods}
\label{sec.method}

\subsection{Task}
Our task is to use the information from above neck body parts to finish complex control tasks on computer. To be specifically, we choose to fully control the first-person 3A game ``Black Myth: WuKong". This requires basic character moving, perspective changing, mouse selection, 27 keys needed to be pressed. Beyond that, it requires real-time processing for users to play well in the game, otherwise it can be very boring due to delay.

\subsection{Overview}
\begin{figure}
    \centering
    \includegraphics[width=1\linewidth]{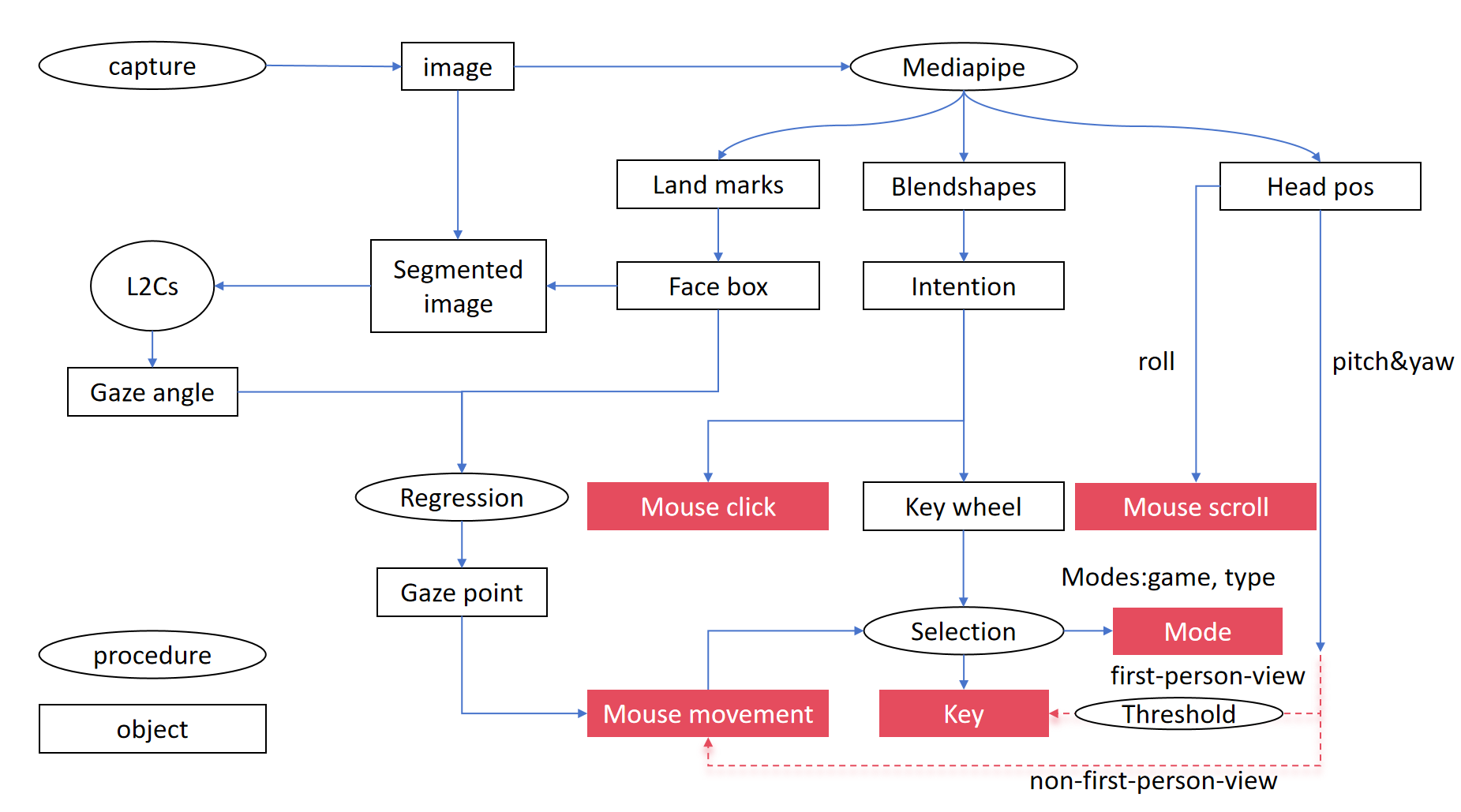}
    \caption{The NeuGaze framework captures a 30 Hz webcam image, processed by Mediapipe for facial landmarks, blend shapes, and head poses. Landmarks define a face box, segmented for the L2CS model to predict gaze angles (yaw, pitch). These, with the face box, feed a LASSOCV regression model to estimate gaze points for mouse control. Blend shapes, thresholded into binary values, combine logically to detect intentions (e.g., jaw movements), mapping to key presses or mouse clicks. Head poses (pitch, yaw, roll) control mouse movement, key selection, or scrolling, with mode switching based on user preference.}
    \label{fig:neugaze}
\end{figure}

Our NeuGaze framework provides a simple, cost-effective solution for individuals with severe motor impairments, enabling hands-free computer interaction using only a standard webcam (30 Hz), which is often pre-installed in laptops, incurring zero additional cost. By leveraging head movements, facial expressions, and eye gaze, NeuGaze achieves robust control outcomes with minimal hardware. This section describes the framework’s components, gaze-based mouse control, intention detection, and head pose utilization, as illustrated in Figure \ref{fig:neugaze}. NeuGaze processes webcam images to extract three signals: gaze angles, facial blend shapes, and head poses. Gaze angles drive mouse cursor movement, blend shapes detect user intentions (e.g., facial expressions) for key presses or clicks, and head poses enable scrolling, mode switching, or alternative cursor control. The system uses Mediapipe \cite{lugaresi2019mediapipe} for facial and head analysis and the L2CS model \cite{L2CS-NET-2023} for gaze estimation, requiring only a single calibration to adapt to individual users. Its design prioritizes simplicity, accessibility, and personalization, avoiding the complexity of traditional brain-computer interfaces (BCIs).

\subsection{System Components}
The NeuGaze pipeline processes a 30 Hz webcam image as follows:
\begin{itemize}
    \item \textbf{Facial Analysis}: Mediapipe \cite{lugaresi2019mediapipe} extracts facial landmarks (468 points), blend shapes (52 continuous values representing expressions, e.g., jawOpen), and head poses (3D vector: pitch, yaw, roll).
    \item \textbf{Gaze Estimation}: Four landmark points form a face box, segmenting the image for the L2CS model \cite{L2CS-NET-2023}, which predicts gaze angles (yaw, pitch, 0–360°). A LASSOCV regression model maps gaze angles and face box coordinates to screen gaze points.
    \item \textbf{Output Mapping}: Gaze points control mouse movement, blend shapes trigger intentions (e.g., clicks), and head poses manage scrolling or mode selection.
\end{itemize}
Calibration involves users staring at predefined screen points while varying head positions, capturing gaze angles and face boxes to train the regression model. LASSOCV was selected over other scikit-learn models (e.g., MLP) for its superior generalization across the screen, avoiding overfitting to sparse training points.

\subsection{Gaze-Based Mouse Control}
NeuGaze supports two mouse control modes for user to use mouse control. 
\begin{itemize}
    \item \textbf{Absolute Mode}: The predicted gaze point directly maps to mouse coordinates.
    \item \textbf{Relative Mode}: If the gaze point enters a screen corner, it triggers continuous mouse movement in that direction, enabling viewpoint changes in gaming or scrolling applications. 
\end{itemize}

The system automatically detects if it is in first-person perspective, if so, it will use relative mode, otherwise it will use absolute mode.

\subsection{Intention Detection via Blend Shapes}
Intentions, defined as distinct facial expressions, are detected by thresholding and combining Mediapipe blend shapes into binary values. For example, the intention “jaw open” is defined as \texttt{jawOpen > 0.4 \& jawLeft < 0.1 \& jawRight < 0.1}, distinguishing it from “jaw left” or “jaw right.” This granularity triples the number of detectable intentions, avoiding conflicts (e.g., misinterpreting a left jaw movement as only “jaw open”). Thresholds are user-specific, accommodating individual facial muscle control, and are set during calibration. Intentions map to actions:
\begin{itemize}
    \item \textbf{Key Wheel}: Each intention (e.g., jaw open) will open the corresponding virtual wheel of keys (e.g., letters, commands).
    \item \textbf{Mouse Clicks}: Intentions trigger left/right clicks, supporting standard or gaming interfaces.
\end{itemize}

\subsection{Head Pose Utilization}
Head poses (pitch, yaw, roll) enhance control versatility:
\begin{itemize}
    \item \textbf{Scrolling}: Roll angles exceeding a threshold (e.g., |roll| > 10°) trigger mouse scrolling, proportional to the angle.
    \item \textbf{Cursor Control}: In non-first-person scenarios, pitch and yaw angles greater than the threshold drive mouse movement, offering an alternative to gaze-based control.
    \item \textbf{Key Selection}: When a key wheel is open, use head poses or gaze to select a key, and undoing the intention confirms the choice.
    % \item \textbf{Mode Switching}: Specific head pose sequences (e.g., pitch up then down) toggle between type and game modes, based on user preference.
\end{itemize}

% This multi-modal approach ensures NeuGaze adapts to diverse applications, from communication to gaming, using only webcam input.

\subsection{Keys mapping}

As I mentioned earlier, there are 27 keys need to be controled, it is extremely difficult for a person to do 27 types of facial expressions precisely. Therefore, we used head motion here to do character moving, and scroll up or down. Besides, we used key wheel to reduce intention needed. We used one intention (only left smile) to map [z,x,c], which are three types of stick style, one intention (closed mouth to left) to map [1,2,3,4], which are 4 skills of WuKong, and one intention (closed mouth to right) to map [q,r,f,t], which are 4 types of buff or tool. For directly control keys, for example, we have intention (only smile right) to do whirl stick, intention (up-lifting eyebrows) for dodge, intention (mouth pucking) for left click and intention (jaw to right) for right click. After all, we have mapped all the facial expressions we can make along with gaze control and head control to all the keys needed in the game, full mapping details are shown in Tab.~\ref{lst:yaml-Snippet}.

NeuGaze leverages a 30 Hz webcam, Mediapipe, and the L2CS model to deliver a low-cost, effective alternative to BCIs. By integrating gaze estimation, facial intention detection, and head pose tracking, it supports mouse control, key selection, and scrolling with minimal hardware. User-specific calibration and thresholding ensure accessibility, making NeuGaze a practical solution for motor-impaired individuals.

\section{Result}

This experiment is carried by myself on my laptop, which has a camera of image size 640x480 and an NVIDIA 4060 GPU. You do not need anything else! Our NeuGaze can run at 30Hz which is the same as camera FPS.

%However, since this paper is not a strictly technical paper, I didn't test these. If you are interested, and think this paper can be published in a good journal or conference with your help, like writing the paper and doing some experiments, please contact me.

% \newpage

The most important thing is that we can really play video games with our head, face and eyes without hands or feet. Here are some screenshots to understand what is going on, how can I control the key and mouse in the game. We can only show screenshots here which is not showing all the sequential information, if you want to see more details, you can watch on \href{https://www.youtube.com/watch?v=s8m4M9XCUSo&t=39s}{YOUTUBE} or \href{https://www.bilibili.com/video/BV1kKdYYVEEM/?spm_id_from=333.1387.homepage.video_card.click&vd_source=80cfd0f23a737ec4dee4be764cccc8fc}{BILIBILI}.

\subsection{First-Person Gaming Controls}
\begin{figure}[htbp]
    \centering
    \begin{minipage}[t]{0.48\linewidth}
        \includegraphics[width=\linewidth]{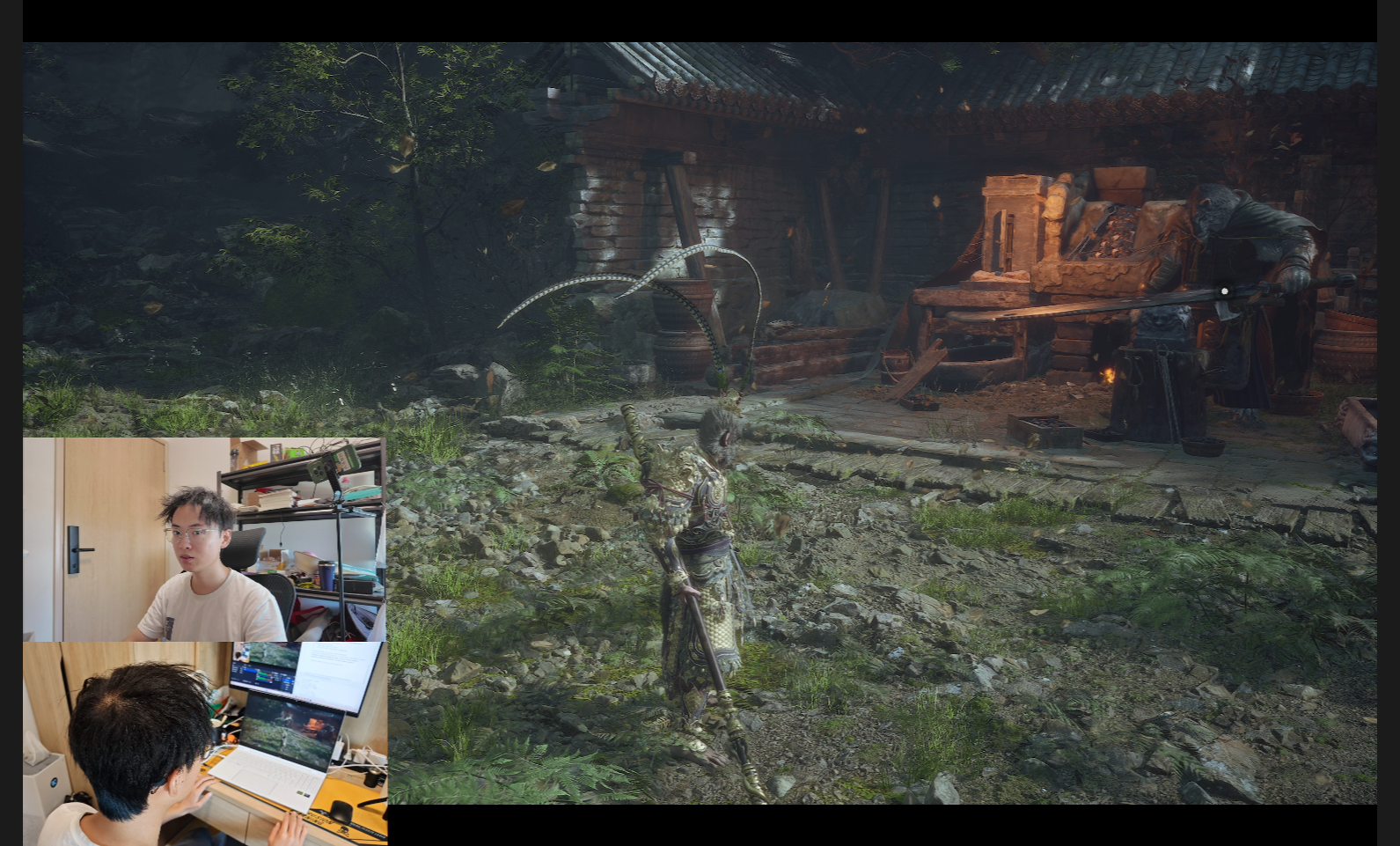}
        \caption{Changing the camera perspective.}
        \label{fig:changing-view}
    \end{minipage}
    \hfill
    \begin{minipage}[t]{0.48\linewidth}
        \includegraphics[width=\linewidth]{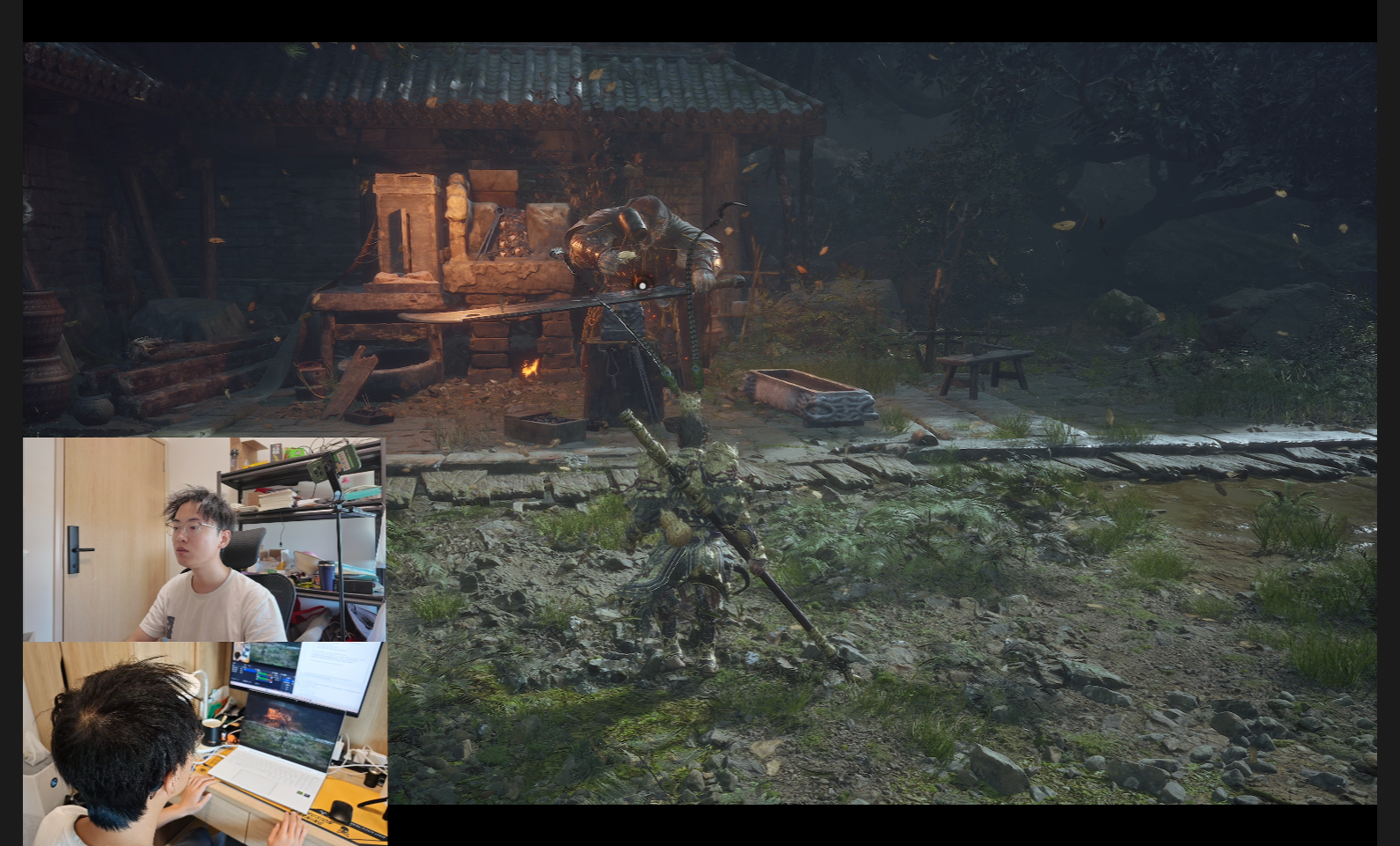}
        \caption{Moving forward the character.}
        \label{fig:moving-forward}
    \end{minipage}
    \caption{Basic in-game controls: adjusting the camera angle and moving through the environment.}
    \label{fig:basic-controls}
\end{figure}

First-person games require precise control of perspective and character movement, which NeuGaze achieves using eye gaze and head motion. When the gaze point approaches the screen’s edges, the viewpoint adjusts accordingly; for example, a gaze point in the left corner shifts the perspective leftward, as shown in Figure \ref{fig:changing-view}. Character movement is controlled by head angles: tilting the head upward (positive pitch) simulates continuous pressing of the ‘W’ key, moving the character forward, as illustrated in Figure \ref{fig:moving-forward}. This approach ensures seamless navigation in dynamic gaming environments.

\subsection{Cursor item selection}
\begin{figure}[htbp]
    \centering
    \begin{minipage}[t]{0.48\linewidth}
        \includegraphics[width=\linewidth]{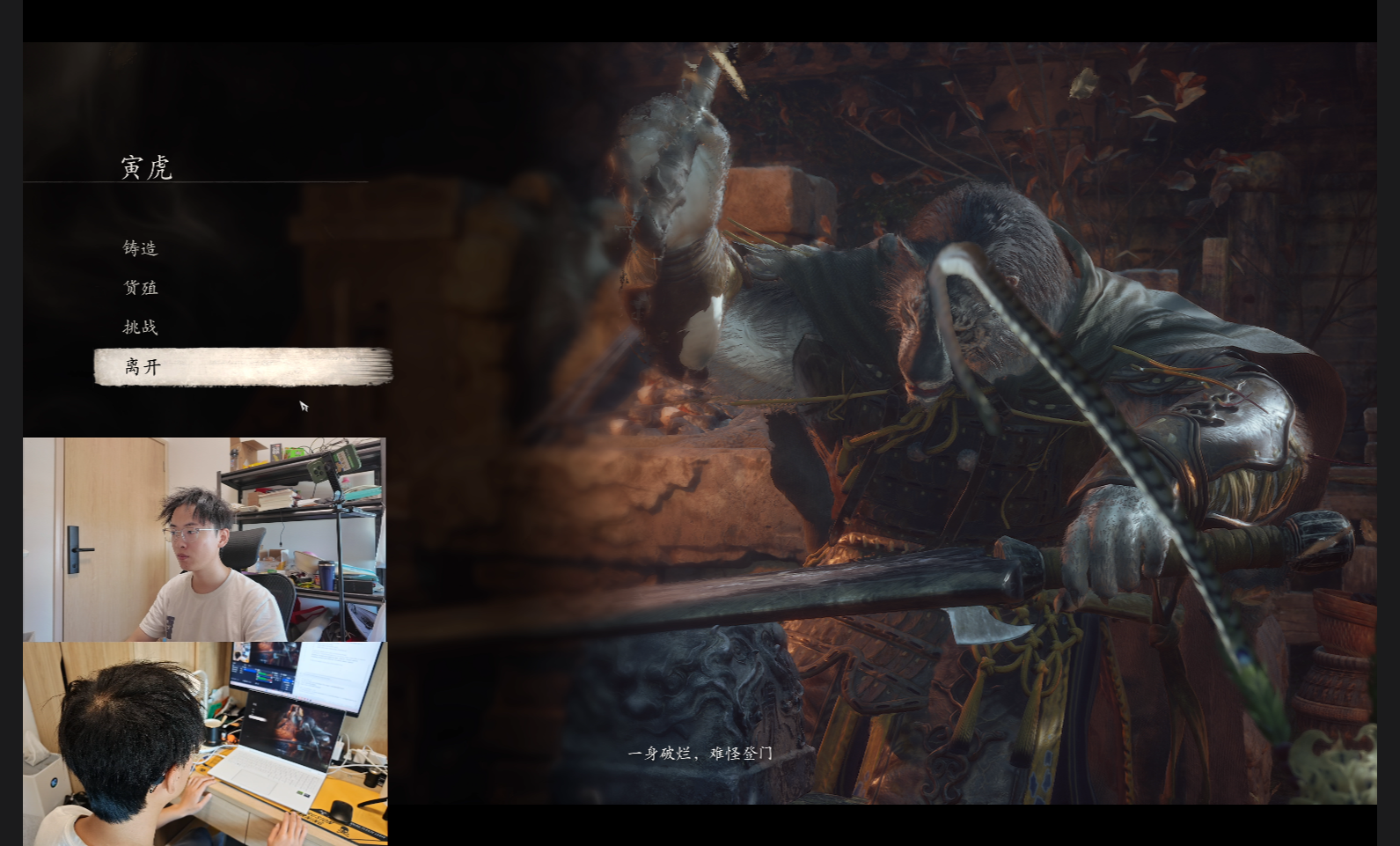}
        \caption{Converting gaze point coordinates to mouse cursor position.}
        \label{fig:gaze-to-mouse}
    \end{minipage}
    \hfill
    \begin{minipage}[t]{0.48\linewidth}
        \includegraphics[width=\linewidth]{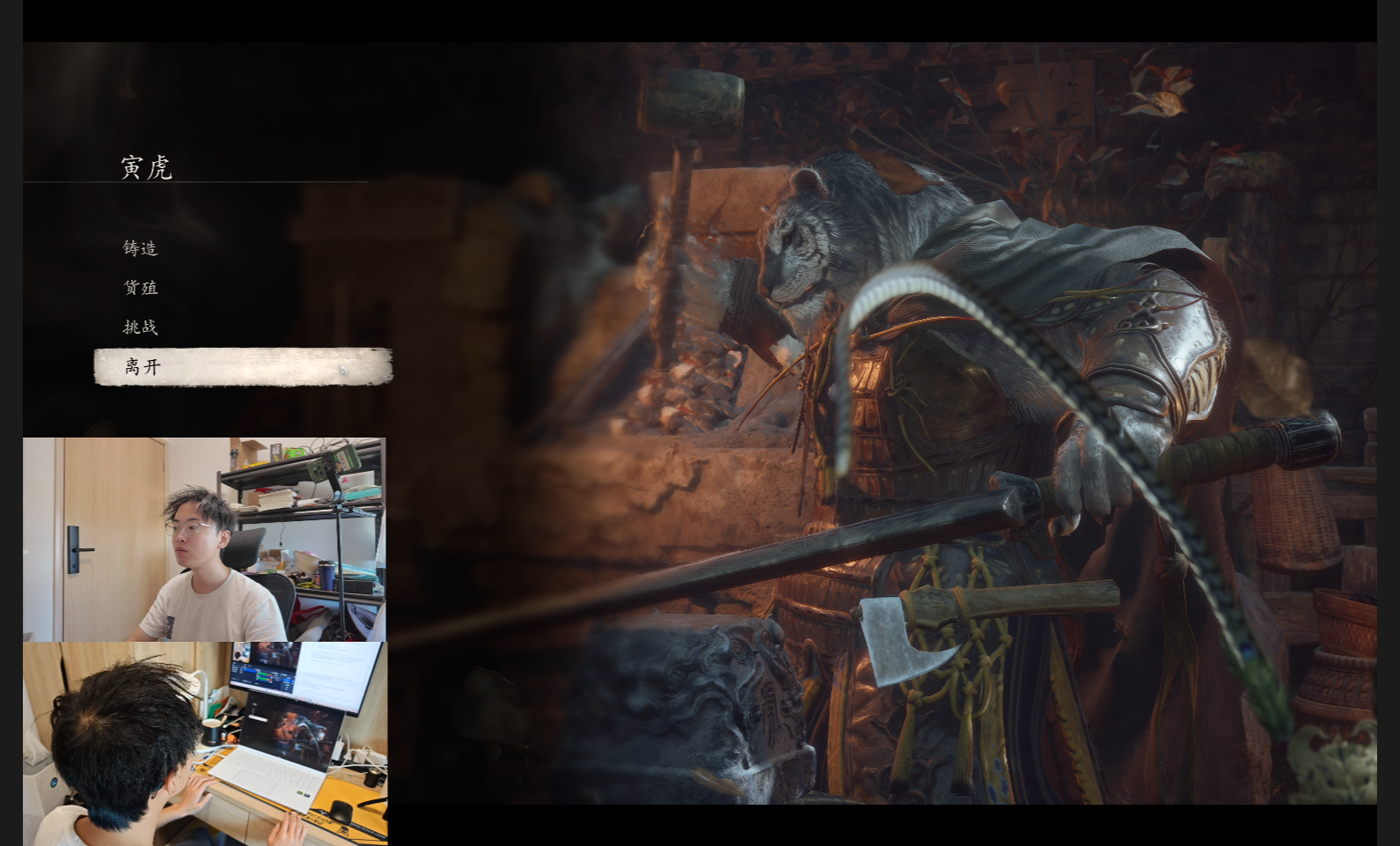}
        \caption{Head movement refines cursor position when gaze tracking is inaccurate.}
        \label{fig:head-mouse}
    \end{minipage}
    
    \vspace{0.5cm} % 增加行间距
    
    \begin{minipage}[t]{0.48\linewidth}
        \includegraphics[width=\linewidth]{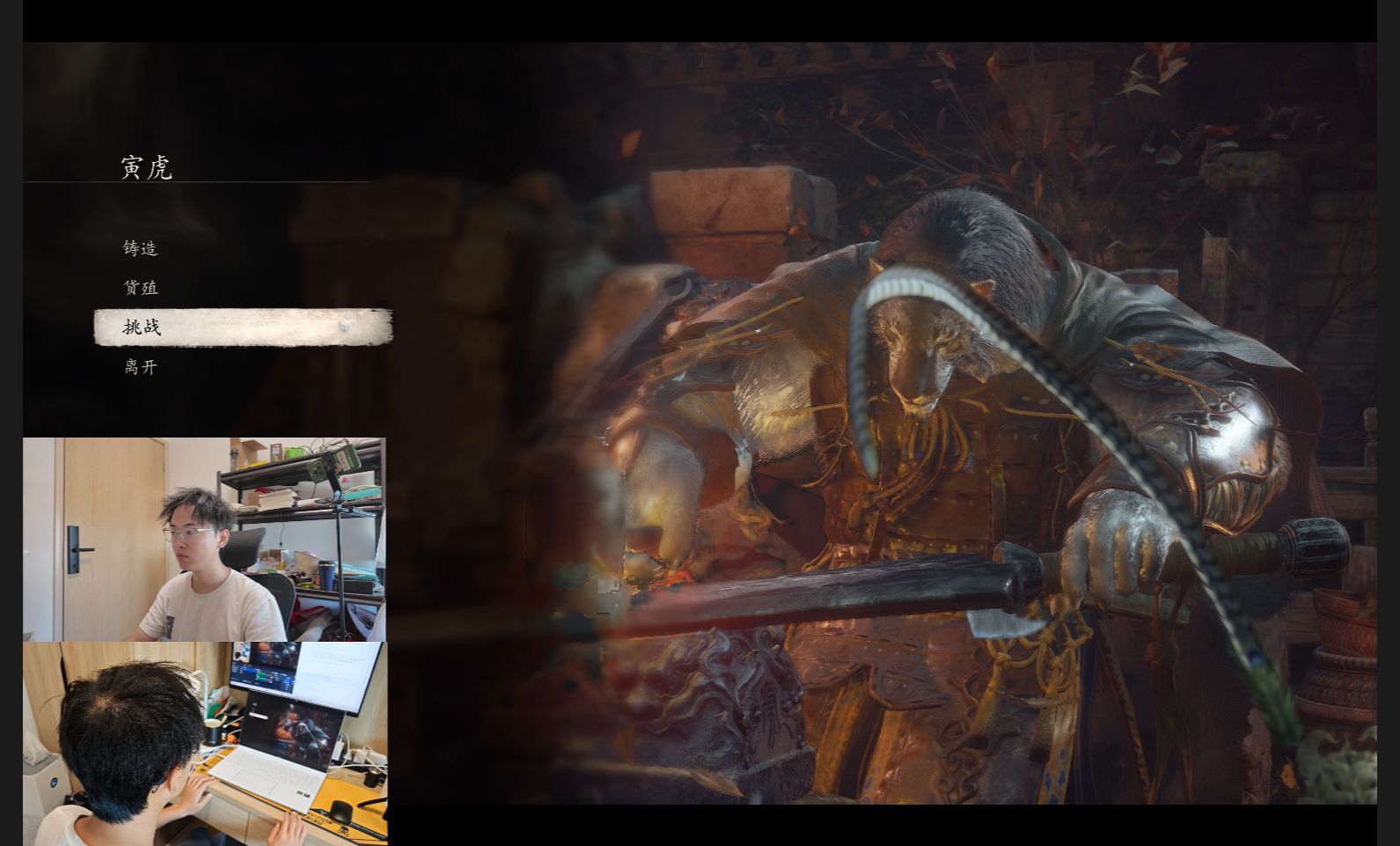}
        \caption{Cursor remains stationary for 1 second after head-based adjustment.}
        \label{fig:mouse-stay}
    \end{minipage}
    \hfill
    \begin{minipage}[t]{0.48\linewidth}
        \includegraphics[width=\linewidth]{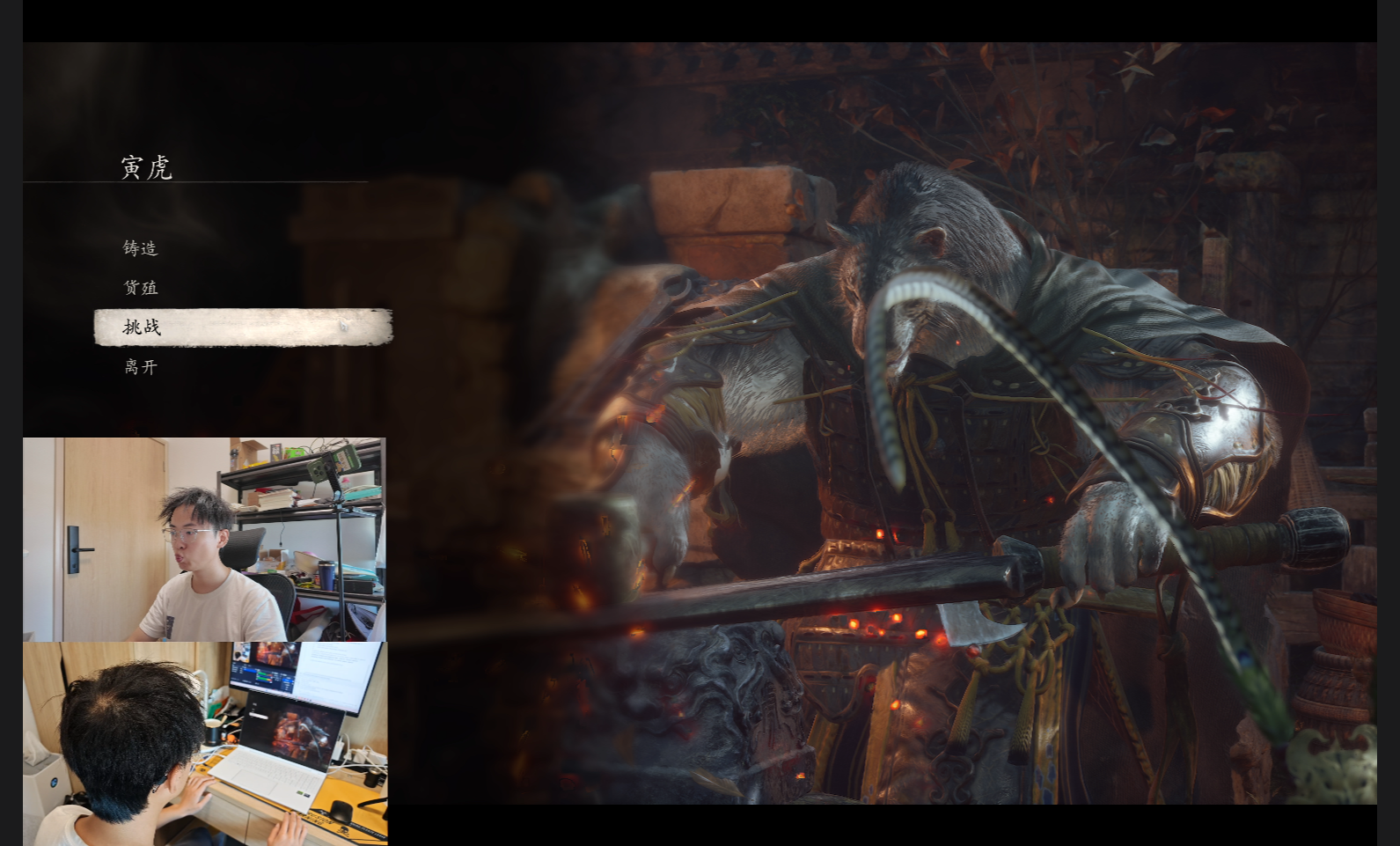}
        \caption{Mouse click triggered by a mouth pucker gesture.}
        \label{fig:mouth-click}
    \end{minipage}
    
    \caption{Interaction workflow combining gaze tracking, head movement, and gesture control.}
    \label{fig:interaction-system}
\end{figure}

Daily computer use involves frequent cursor movement and selection, which NeuGaze streamlines through a hybrid gaze-head approach, as depicted in Figure \ref{fig:interaction-system}. Initially, the gaze point rapidly positions the cursor near a target region (Figure \ref{fig:gaze-to-mouse}). Since gaze estimation may slightly deviate from the intended point, head movements (pitch and yaw) enable fine adjustments to the cursor’s x- and y-coordinates for precise selection (Figure \ref{fig:head-mouse}). The cursor then pauses for 1 second to allow actions such as clicking, double-clicking, or right-clicking (Figure \ref{fig:mouse-stay}). Finally, a mouth pucker expression triggers a mouse click (Figure \ref{fig:mouth-click}), ensuring efficient and accurate interactions.

\subsection{Key Triggering Mechanisms}
In gaming, specific keys like ‘E’ (interaction) and ‘SPACE’ (dodge) are critical. NeuGaze assigns facial expressions based on their temporal suitability. A pressed mouth expression, less prone to rapid repetition, triggers the ‘E’ key for sustained interactions, as shown in Figure \ref{fig:directly_trigger}. Conversely, raising eyebrows, easily performed quickly and repeatedly, activates the ‘SPACE’ key for dodging. Additional actions, such as opening a backpack or executing a whirl, can be mapped to other expressions, enhancing the system’s versatility.

\subsection{Wheel for Efficient Control}
\begin{figure}[htbp]
    \centering
    \begin{minipage}[b]{0.48\linewidth}
        \includegraphics[width=\linewidth]{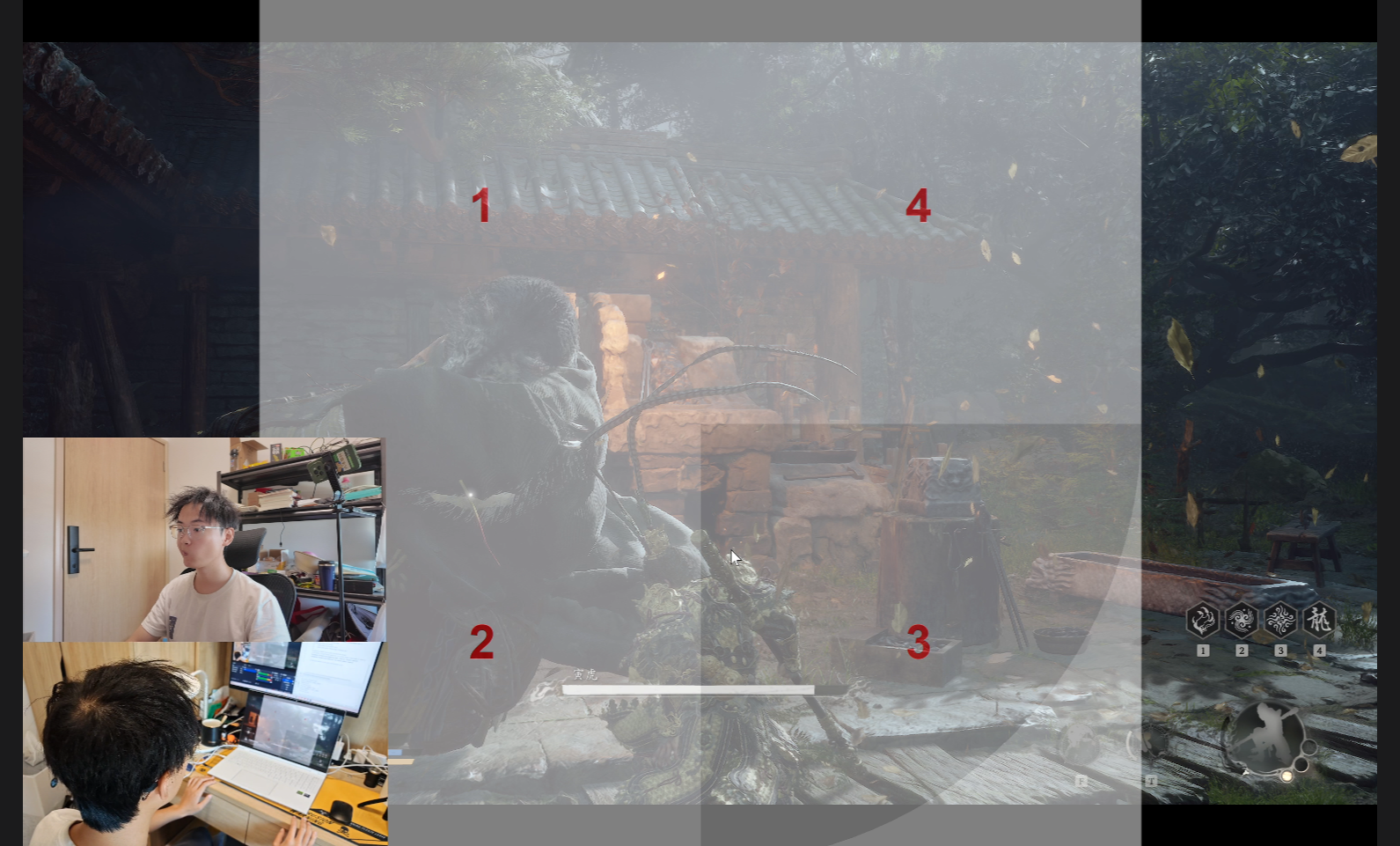}
        \caption{Trigger Skill 2}
        \label{fig:trigger-skill2}
    \end{minipage}
    \hfill
    \begin{minipage}[b]{0.48\linewidth}
        \includegraphics[width=\linewidth]{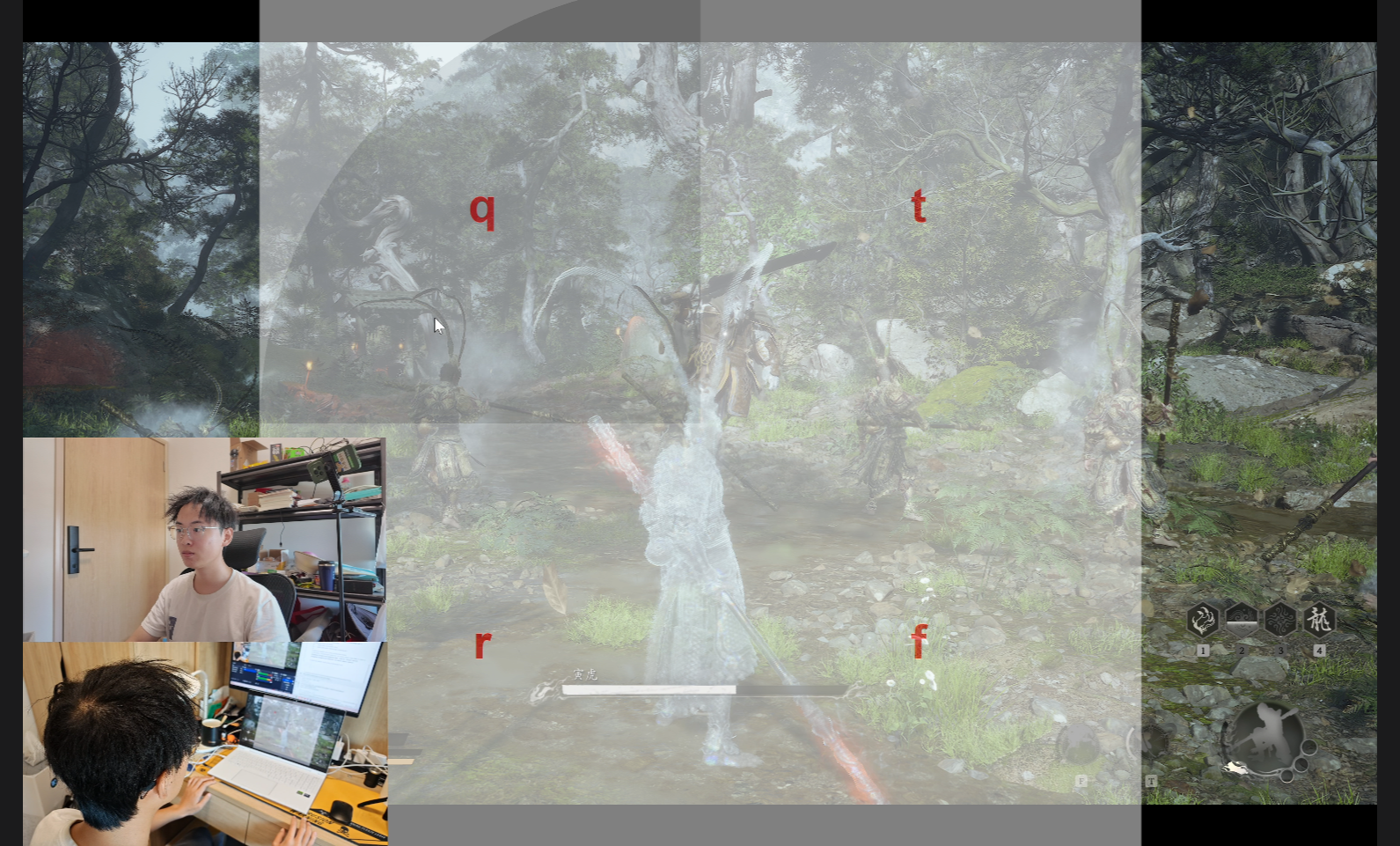}
        \caption{Trigger Q Eat Drug}
        \label{fig:trigger-q-eat-drug}
    \end{minipage}
    \caption{The concept of selection wheel. We can save a lot of facial intentions since many can be integrated into one intention, instead of using each intention to trigger each key.}
    \label{fig:trigger-selection-wheel}
\end{figure}

Assigning a unique facial expression to each game key is impractical due to the limited number of distinct expressions users can reliably produce. NeuGaze introduces a concept called wheel, displayed on-screen, to select keys efficiently (Figure \ref{fig:trigger-selection-wheel}). For instance, to access four in-game skills, a leftward mouth movement activates the skill wheel; the user selects a skill using eye gaze or head angles (pitch, yaw) and returns the mouth to neutral to confirm (Figure \ref{fig:trigger-skill2}). Similarly, actions like consuming a drug (key ‘Q’) are selected via the wheel (Figure \ref{fig:trigger-q-eat-drug}). This reduces the required facial expressions from eight to two for controlling eight keys, lowering complexity and enhancing usability.

\subsection{Demonstration: Defeating the Yin Tiger}
\begin{figure}[htbp]
    \centering
    \begin{minipage}[b]{0.48\linewidth}
        \centering
        \includegraphics[width=\linewidth]{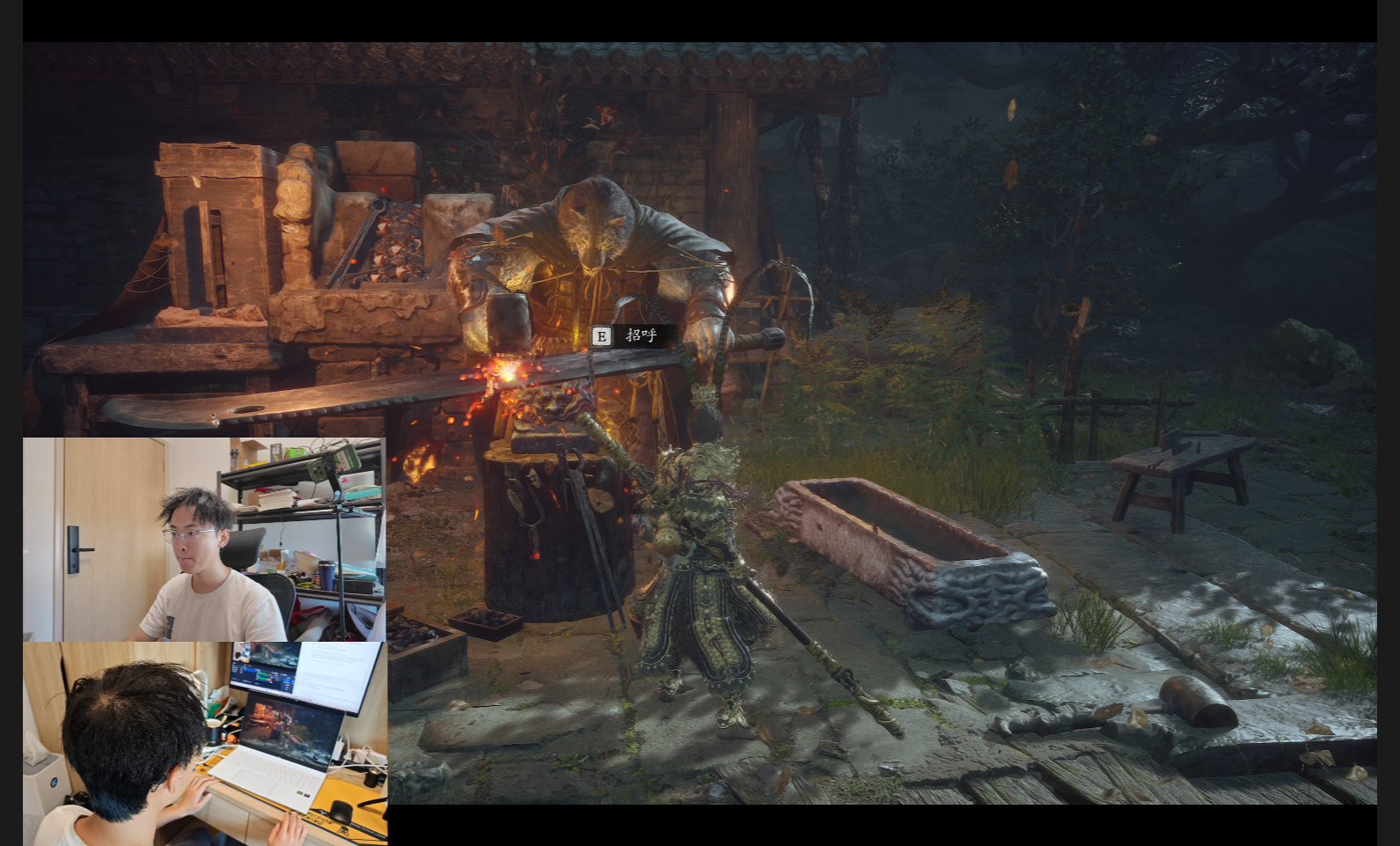}
        \caption*{(a) Trigger E with mouth pressed.}
        \label{fig:trigger_E}
    \end{minipage}
    \hfill
    \begin{minipage}[b]{0.48\linewidth}
        \centering
        \includegraphics[width=\linewidth]{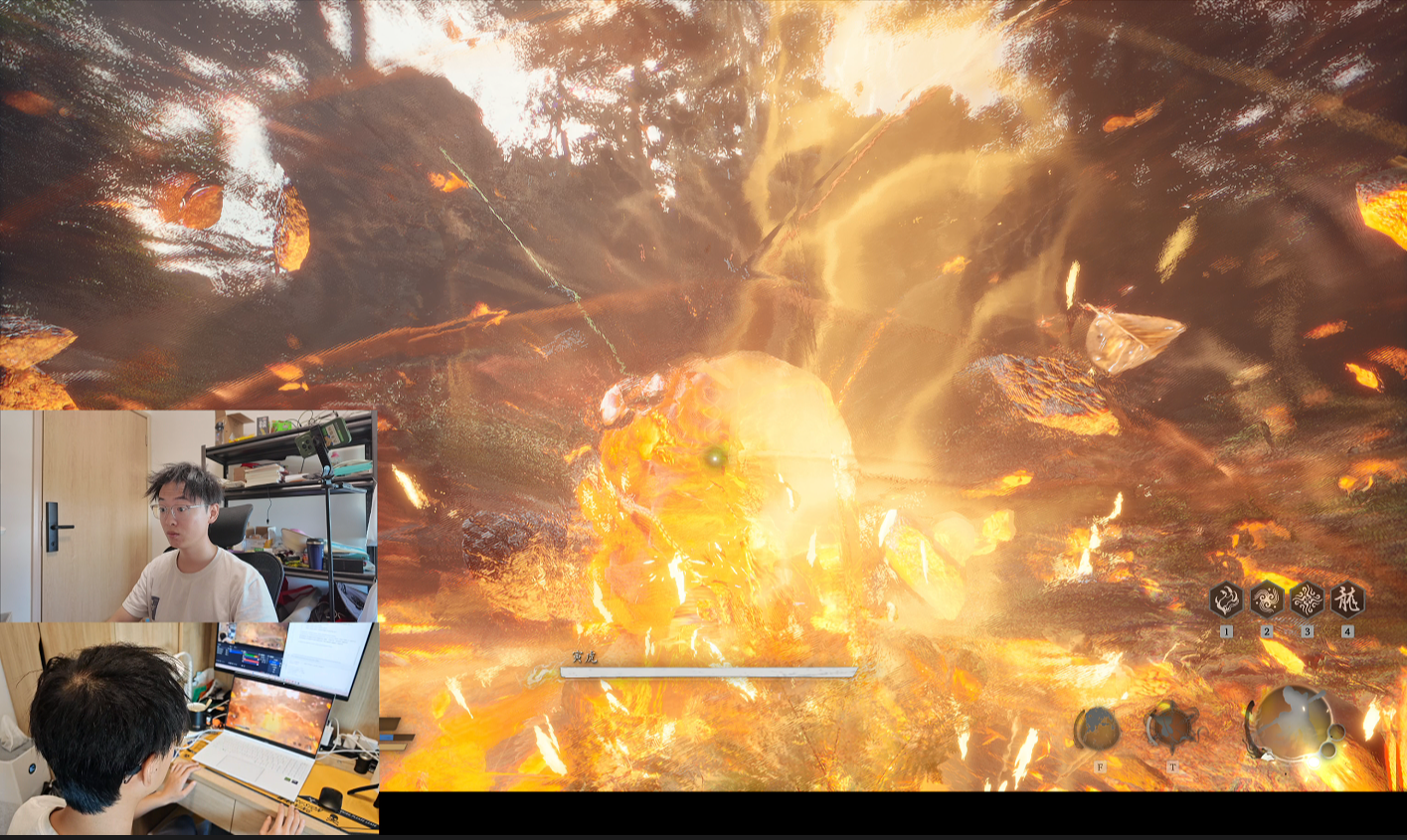}
        \caption*{(b) Dodge using lifting eyebrows.}
        \label{fig:dodge-lift-eyebrows}
    \end{minipage}
    
    \begin{minipage}[b]{0.48\linewidth}
        \includegraphics[width=\linewidth]{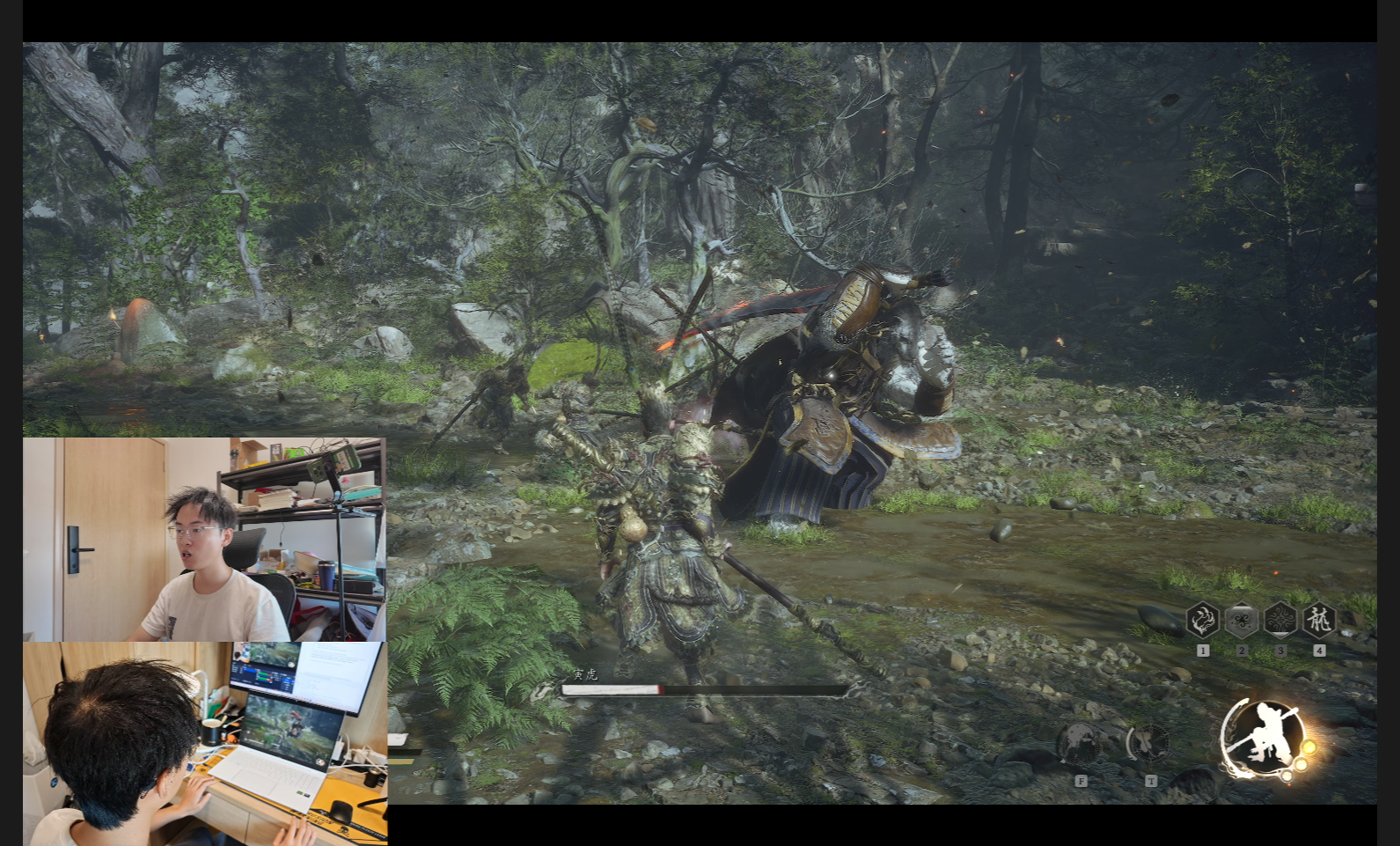}
        \caption{Heavy Attack}
        \label{fig:heavy-attack}
    \end{minipage}
    \hfill
    \begin{minipage}[b]{0.48\linewidth}
        \includegraphics[width=\linewidth]{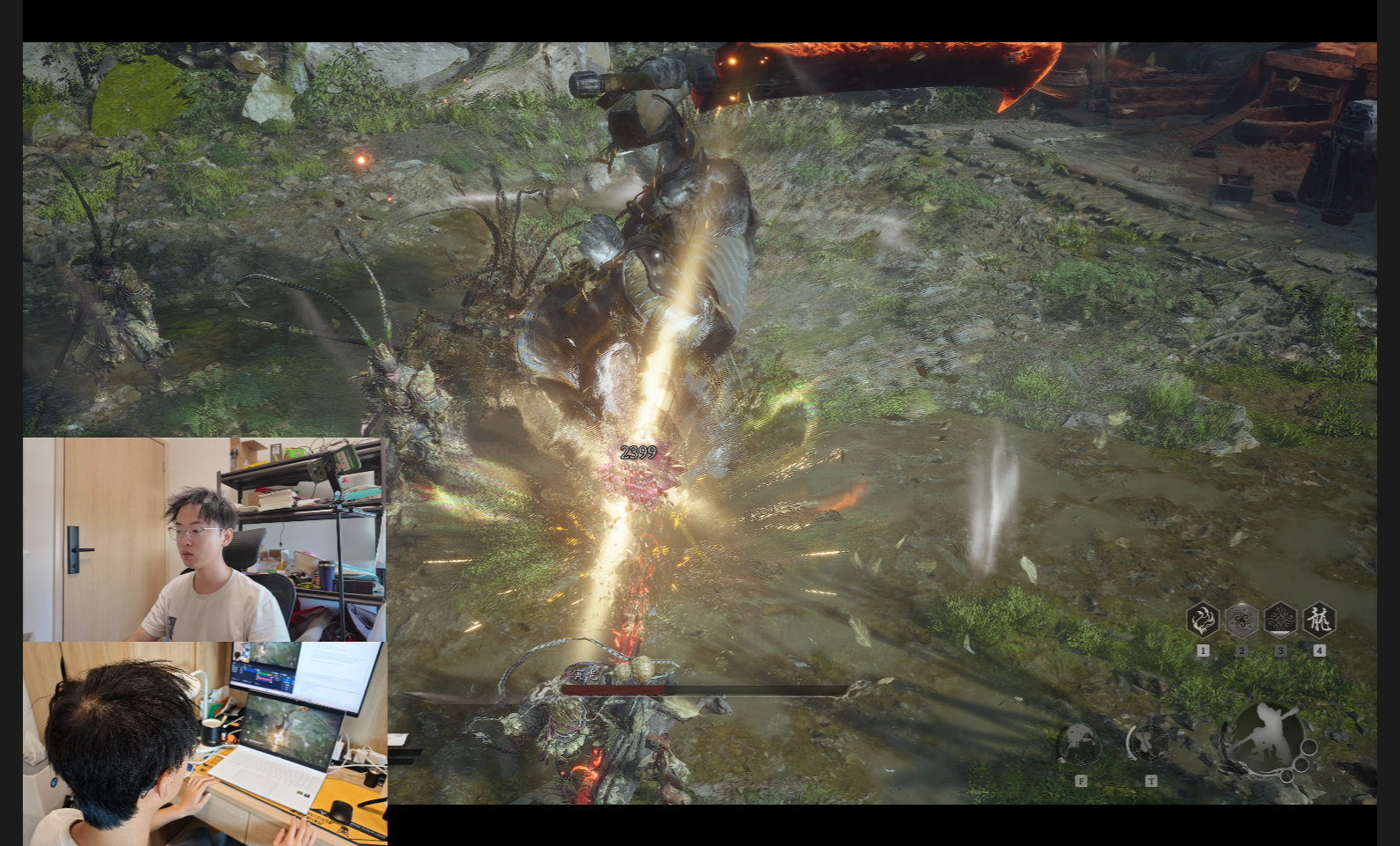}
        \caption{Defeat Tiger with Heavy Attack}
        \label{fig:defeat-tiger}
    \end{minipage}
    \caption{Three examples of trigger keys directly.}
    \label{fig:directly_trigger}
\end{figure}

Using these controls, NeuGaze enables complex gaming interactions, demonstrated by defeating the Yin tiger, a formidable boss monster, as shown in Figure \ref{fig:defeat-tiger}. This success underscores the framework’s ability to integrate gaze, head, and facial controls for precise and responsive game play.

% \begin{figure}
%     \centering
%     \includegraphics[width=1\linewidth]{figures/trigger_E.png}
%     \caption{Trigger E}
%     \label{fig:trigger_E}
% \end{figure}
% \begin{figure}
%     \centering
%     \includegraphics[width=1\linewidth]{figures/dodge.png}
%     \caption{Dodge using lifting eyebrows.}
%     \label{fig:dodge-lift-eyebrows}
% \end{figure}

% \begin{figure}[htbp]
%     \centering
%     \begin{minipage}[b]{0.48\linewidth}
%         \includegraphics[width=\linewidth]{figures/heavy_attack.png}
%         \caption{Heavy Attack}
%         \label{fig:heavy-attack}
%     \end{minipage}
%     \hfill
%     \begin{minipage}[b]{0.48\linewidth}
%         \includegraphics[width=\linewidth]{figures/defeat_tiger_with_heavy_attack.png}
%         \caption{Defeat Tiger with Heavy Attack}
%         \label{fig:defeat-tiger}
%     \end{minipage}
%     \caption{Trigger heavy attack. The heavy attack has internal delay in the game, so we have to put two screenshots to show my intention and effect. }
%     \label{fig:attack-comparison}
% \end{figure}

\FloatBarrier

\section{Limitations}

While NeuGaze demonstrates promising performance in enabling hands-free control for complex tasks, several limitations warrant consideration. First, the current evaluation lacks comprehensive quantitative analysis. Testing was conducted by a single user on a specific hardware setup, limiting the generalizability of performance metrics such as accuracy, latency, or user satisfaction. Future studies should incorporate multi-user experiments to quantify these metrics and validate system robustness across diverse populations.

Second, the user experience is not consistently seamless. The system’s reliance on a 30 Hz webcam and real-time processing occasionally results in suboptimal responsiveness, particularly under varying lighting conditions or with rapid facial movements. Engaging a broader user base, including iterative feedback from motor-impaired individuals, is essential to refine the system’s usability and ensure a smooth interaction experience.

Third, the absence of testing with motor-impaired patients represents a significant gap. NeuGaze’s design targets populations with conditions like quadriplegia or ALS, yet logistical constraints prevented their participation in this study. Involving these users in future iterations will be critical to identify specific usability challenges, such as calibration difficulties or fatigue, and to optimize the system for their needs.

Finally, the current key mapping design lacks universal applicability. Tailored specifically for \textit{Black Myth: Wukong} and calibrated to the facial muscle capabilities of a single user, the mappings are not readily adaptable to other games or users with varying motor abilities. Unlike traditional keyboard and mouse inputs, which offer broad compatibility, NeuGaze requires customized configurations, limiting its immediate versatility. Future work should explore standardized mapping frameworks to enhance generalizability across applications and user profiles.

Addressing these limitations through expanded testing, user-inclusive design, and adaptive key mapping will strengthen NeuGaze’s potential as a transformative assistive technology.

\section{Conclusion}

% 这一切都是从重新发掘脑机接口适用人群的现状和需求出发的，而不是从现有的脑机接口的技术上做的增量创新。换而言之，这是从第一性原理出发的突破性创新。
% 我们将头部，面部，眼部的信息以一种新的方式组合利用起来。在玩黑神话悟空这样复杂的需要实时控制的第一人称视角的动作游戏中实现了基础的视角移动和位置移动操作，鼠标移动选择点击操作，触发按键操作，引入了wheel的概念来实现用极少的表情来控制最多的按键，并且最终打败了重要的BOSS 怪，说明了我们NeuGaze系统的实用性。

% 我认为NeuGaze这种控制方式将会在未来几年深入渗透到脑机接口的应用场景中，用低成本、安全、有效的特性为广大的脑机接口适用群体服务，
NeuGaze introduces a groundbreaking webcam-based paradigm for human-computer interaction, leveraging eye gaze, head movements, and facial expressions to achieve BCI-like functionality without specialized hardware. By enabling real-time control of a complex game, \textit{Black Myth: Wukong}, NeuGaze demonstrates performance comparable to conventional inputs, with the skill wheel reducing the cognitive load of key triggering. Its success in defeating a formidable boss underscores its potential for motor-impaired users. However, current results are limited to a single user, and environmental factors like lighting may affect performance. Future work includes multi-user studies, quantitative comparisons with existing BCIs, and exploration of non-gaming applications, such as communication or smart home control. By prioritizing accessibility and simplicity, NeuGaze paves the way for cost-effective, inclusive assistive technologies, redefining the landscape of human-computer interaction.

% This research began by re-evaluating the current needs and conditions of brain-computer interface (BCI) users, rather than pursuing incremental improvements based on existing BCI technologies. In other words, it represents a breakthrough innovation grounded in first principles.

% We have integrated information from the head, face, and eyes in a novel way. In complex, real-time control scenarios like the first-person action game Black Myth: Wukong, our NeuGaze system achieved fundamental operations such as viewpoint and positional movement, mouse-based selection and clicking, and key trigger actions. By introducing the concept of a "wheel," we enabled a minimal set of facial expressions to control a wide range of key inputs. The successful defeat of a significant boss in the game demonstrates the practical utility of the NeuGaze system.

% We believe that the NeuGaze control paradigm will deeply penetrate BCI application scenarios in the coming years, offering low-cost, safe, and effective solutions to serve a broad range of BCI users. By prioritizing accessibility and user-centric design, NeuGaze has the potential to redefine the landscape of BCI technology, empowering individuals with diverse needs to interact seamlessly with digital environments and unlocking new possibilities for human-computer interaction.
% \FloatBarrier
\clearpage

\medskip

{
\small
\bibliography{IEEEabrv, references}
}

\newpage
\appendix

\begin{lstlisting}[style=yaml, caption={YAML Config Snippet}, label={lst:yaml-Snippet}]

head_angles_center:
  yaw: 0
  pitch: 3
  roll: 0

head_angles_scale:
  yaw: 8
  pitch: 8
  roll: 8

key_config: 
  game:
    numlock:
      wheel: [game, type ]
    num0:
      wheel: [e]
    num1:
      wheel: [z, x, c]
    num2:
      wheel: [shift]
    num3:
      wheel: [v]
    num4:
      wheel: [1, 2, 3, 4]
    num5:
      wheel: [g]
    num6:
      wheel: [q, r, f, t]
    num7:
      wheel: [esc]
    num8:
      wheel: [space]
    num9:
      wheel: [ctrl]
    left_click:
      wheel: [mouse_left]
      induce:
        lock_mouse_move:
          duration: 1
    mid_click:
      wheel: [mouse_middle]
    right_click:
      wheel: [mouse_right]
    extra:
      wheel: [null]
    head_up:
      wheel: [s]
    head_down:
      wheel: [w]
    head_left:
      wheel: [a]
    head_right:
      wheel: [d]
    head_roll_left:
      wheel: [scroll_up]
    head_roll_right:
      wheel: [scroll_down]

  type:
    numlock:
      wheel: [game, type ]
    num0:
      wheel: [keydown, keyup]
    num1:
      wheel: [backspace]
    num2:
      wheel: [shift, ctrl, caps, tab, alt, esc, fn, win]
    num3:
      wheel: [ctrl+c, ctrl+v, ctrl+q, ctrl+a, ctrl+alt]
    num4:
      wheel: [a, b, c, d, e, f, g, h, i, j, k, l, m, n, o, p, q, r, s, t, u, v, w, x, y, z, backspace, space, enter]
      layout_type: square
    num5:
      wheel: [null]
    num6:
      wheel: ['0', '1', '2', '3', '4', '5', '6', '7', '8', '9', '`', '-', '=', '[', ']', '\', ';', "'", ',', '.', '/']
      layout_type: square
    num7:
      wheel: [esc]
    num8:
      wheel: [space]
    num9:
      wheel: [F1, F2, F3, F4, F5, F6, F7, F8, F9, F10, F11, F12]
    left_click:
      wheel: [mouse_left]
      induce:
        lock_mouse_move:
          duration: 1
    mid_click:
      wheel: [mouse_middle]
    right_click:
      wheel: [mouse_right]
    extra:
      wheel: [null]
    head_roll_left:
      wheel: [scroll_up]
    head_roll_right:
      wheel: [scroll_down]


expression_evaluator_config:
  expressions:
    numlock:
      conditions:
        - feature: jawOpen
          operator: ">"
          threshold: 0.4
        - feature: jawLeft
          operator: "<"
          threshold: 0.1
        - feature: jawRight
          operator: "<"
          threshold: 0.1
      combine: "AND"

    num0:
      conditions:
        - feature: mouthRollLower
          operator: ">"
          threshold: 0.45
        - feature: mouthRollUpper
          operator: ">"
          threshold: 0.45
      combine: "AND"

    num1:
      conditions:
        - feature: mouthSmileLeft
          operator: "BETWEEN"
          min: 0.25
          max: 0.45
        - feature: mouthSmileLeft
          operator: "DIFF>"
          compare_to: mouthSmileRight
          threshold: 0.15
      combine: "AND"

    num2:
      conditions:
        - feature: mouthSmileLeft
          operator: ">"
          threshold: 0.45
        - feature: mouthSmileRight
          operator: ">"
          threshold: 0.45
        - feature: mouthSmileLeft
          operator: "DIFF<"
          compare_to: mouthSmileRight
          threshold: 0.2
      combine: "AND"

    num3:
      conditions:
        - feature: mouthSmileRight
          operator: "BETWEEN"
          min: 0.25
          max: 0.45
        - feature: mouthSmileRight
          operator: "DIFF>"
          compare_to: mouthSmileLeft
          threshold: 0.15
      combine: "AND"

    num4:
      conditions:
        - feature: mouthLeft
          operator: ">"
          threshold: 0.2
        - feature: jawOpen
          operator: "<"
          threshold: 0.05
        - feature: mouthSmileLeft
          operator: "<"
          threshold: 0.2
        - feature: mouthSmileRight
          operator: "<"
          threshold: 0.2
      combine: "AND"

    num5:
      conditions:
        - feature: mouthPressLeft
          operator: ">"
          threshold: 0.4
        - feature: mouthPressRight
          operator: ">"
          threshold: 0.4
      combine: "AND"

    num6:
      conditions:
        - feature: mouthRight
          operator: ">"
          threshold: 0.2
        - feature: jawOpen
          operator: "<"
          threshold: 0.05
        - feature: mouthSmileLeft
          operator: "<"
          threshold: 0.2
        - feature: mouthSmileRight
          operator: "<"
          threshold: 0.2
      combine: "AND"

    num7:
      conditions:
        - feature: mouthUpperUpLeft
          operator: ">"
          threshold: 0.5
        - feature: mouthUpperUpRight
          operator: ">"
          threshold: 0.5
        - feature: mouthLowerDownLeft
          operator: ">"
          threshold: 0.3
        - feature: mouthLowerDownRight
          operator: ">"
          threshold: 0.3
      combine: "AND"

    num8:
      conditions:
        - feature: browInnerUp
          operator: ">"
          threshold: 0.8
      combine: "AND"

    num9:
      conditions:
        - feature: mouthFunnel
          operator: ">"
          threshold: 0.4
      combine: "AND"

    right_click:
      conditions:
        - feature: jawLeft
          operator: ">"
          threshold: 0.3
      combine: "AND"

    mid_click:
      conditions:
        - feature: jawRight
          operator: ">"
          threshold: 0.3
      combine: "AND"

    left_click:
      conditions:
        - feature: mouthPucker
          operator: ">"
          threshold: 0.97
        - feature: mouthFunnel
          operator: "<"
          threshold: 0.2
      combine: "AND"

    extra:
      conditions:
        - feature: eyeBlinkLeft
          operator: ">"
          threshold: 0.6
        - feature: eyeBlinkRight
          operator: "<"
          threshold: 0.25
      combine: "AND"

  priority_rules:
    - when: num7
      disable: [num2]
    - when: any
      disable: [left_click]
      except: [left_click]
\end{lstlisting}

\end{document}